\documentclass[a4paper,11pt]{article}
\pdfoutput=1 % if your are submitting a pdflatex (i.e. if you have
\usepackage{jinstpub} % for details on the use of the package, please
\title{\boldmath Beam tests of silicon pixel 3D-sensors developed at SINTEF}

\author[a]{O. Dorholt,}
\author[b]{T. E. Hansen,}
\author[c,1]{A. Heggelund,\note{Now at Department of Physics, University of Oslo, Norway}}
\author[d]{A. Kok,}
\author[c,2]{N. Pacifico,\note{ Now at CERN, Geneva, Switzerland}}
\author[a]{O. Rohne,}
\author[a]{H. Sandaker,}
\author[c]{B. Stugu,}
\author[c]{Z. Yang,}
\author[e]{M. Bomben,}
\author[f,2]{A. Rummler}
\author[g,3]{ and J. Weingarten\note{Now at Lehrstuhl Experimentelle Physik IV, Technische Universit\"at, Dortmund, Germany}}
\affiliation[a]{Department of Physics, University of Oslo, PB 1048 Blindern, 0316 Oslo, Norway}
\affiliation[b] {Hansrad Sensor, 1515 Moss, Norway}
\affiliation[c]{Department of Physics and Technology, University of Bergen,5007 Bergen, Norway}
\affiliation[d]{Department of Microsystems and Nanotechnology (MiNaLab), SINTEF Digital,  0373 Oslo, Norway}
\affiliation[e]{Laboratoire de Physique Nucl\'eaire et de Hautes Energies, (LPNHE) Paris, France}
\affiliation[f]{Laboratoire d'Annecy de Physique des Particules,  74941 Annecy, France}
\affiliation[g]{II. Physikalisches Institut, Georg-August-Universit\"at G\"ottingen, Germany}
\emailAdd{bjarne.stugu@uib.no}

\abstract{For the purpose of withstanding very high radiation doses, silicon pixel sensors with a '3D' electrode geometry are being developed. Detectors of this kind are highly interesting for harch radiation environments such as expected in the High Luminosity LHC, but also for space physics and medical applications. In this paper, prototype sensors developed at SINTEF are presented and results from tests in a pion beam at CERN are given. These tests show that these 3D sensors perform as expected with full efficiency at bias voltages between 5 and 15V. }


\begin{document}
\maketitle
\flushbottom

\section{Introduction}
\label{sec:intro}
The radiation hardness and rate requirements 
for tracking and vertex detectors to be operated in the High luminosity LHC are extremely severe, and a significant R\& D effort is required before a viable system can be realized. The favored technology for the inner tracking elements are silicon pixel detectors. These posess many of the wanted properties and are now routinely used in 
virtually all experiments within high energy physics. However, significant development must be invested into improving the
radiation hardness of the systems. The most exposed areas, close to the colliding beams, must be designed to withstand a fluence 
of $1.4 \times 10^{16}  \rm{cm^{-2}}$ 1 MeV $n_ {eq}$ (neutron equivalents).  It is unlikely that conventional pixel sensors with
planar p-n junctions and standard thickness can tolerate such large fluences. 
Therefore, research on sensors with  alternative designs are being pursued. One 
research path is to drastically modify the diode 
structures. Instead of p-n junctions parallel to the
sensor plane, a proposition where holes (perpendicular to the sensor plane) are etched into the silicon bulk 
and filled with alternate rows of $\rm{p^+}$ and $\rm{n^+}$ type electrodes ~\cite{intro:1} has been successfully 
prototyped ~\cite{intro:2}, and also fabricated at SINTEF MiNaLab{\footnote{Department of Microsystems and Nanotechnology (MiNaLab), SINTEF Digital,  0373 Oslo, Norway} ~\cite{intro:3}.  The advantage of such structures 
is that  full depletion can be achieved at very low bias due to the short distance between the electrodes.  The carrier drift distances are also correspondingly reduced,
resulting in reduced signal loss  due to charge trapping, an effect
that increase in severity as the radiation dose increases.

Very similar designs have also been realised in double sided processes, ~\cite{intro:4,intro:5} and are already successfully operated in the ATLAS IBL, as well as in the ATLAS AFP ~\cite{intro:6}. 
The purpose of this note is to report on developments and tests of  prototypes that have been designed and produced in a single side, active edge process at SINTEF. 
\section{Sensor and module design}
 The sensors tested here are 230 $\mu$m thick and are produced in a single sided process. The pixel size is $ \Delta x \times \Delta y = 250 \times 50 \mu{\rm m^2}$ and is adapted to be read out by the ATLAS Front End readout chip I4 (FEI4) ~\cite{Sensor:1}. The etched vertical electrodes  have a diameter of 14 $\mu $m nominally and are placed 125 $\mu $m apart in rows. The distance along  $y$  between two rows
 is 25 $\mu $m, with a staggered position of electrodes in different rows, resulting in a spacing between electrodes of      
 about 67 $\mu$m.
 \par
 The sensors consists of 80 columns and 336 rows, corresponding to an
 active area of nominally $20 \times 16.8 {\rm mm^2}$ for $250 \times 50 \mu{\rm m^2}$ pixels. A trench
 was made around the nominal sensor area, 16  $\mu$m wide, and centered 72.5 $\mu$m beyond the nominal edge of the outermost pixels, enlarging the sensitive area of the edge pixels in both directions (see sect. 4 for an evaluation of the extended sensitive area.) To be added to this are physical edges. The sensors have been cut using conventional wafer dicing, adding a dead margin of about 100 $\mu$m. The overall physical size of the diced sensor is 20.32 by 17.65 mm$^2$.   
The sensors were bump-bonded to the FEI4 readout chips. This procedure turned out to have some difficulties, and only very few assemblies remained suitable for testing in a test beam. Of the three assemblies put in the testbeam, one suffered from extremely high leakage currents. All three assemblies had dead (or extremely low efficiency) regions due to unsuccesful bump-bonding. However, it was easy to identify these regions, and exclude them from the analysis described below, where the purpose is to test the properties of the sensors, rather than the full module assembly. 
\subsection{Leakage current}
The leakage currents of the modules were measured prior to mounting in 
the test beam, and monitored during the tests. Fig \ref{fig:iv} shows the current characteristics of two  sensor modules  tested. Soft breakdown was observed to be around 12V and 20V. The low values for the breakdown voltages has been understood as pinholes created in metal masks during the DRIE process, as explained in ref.~\cite{Leakage:Laura}. \footnote{Preliminary wafer tests of the subsequent production show that the a solution is found to this problem} The third module had a problem with some large ohmic currents, which is thought to be external to the sensor itself. The module was nevertheless included for studies in the test beam.   
\begin{figure}[htbp]
\centering % \begin{center}/\end{center} takes some additional vertical space
\includegraphics[width=.6\textwidth]{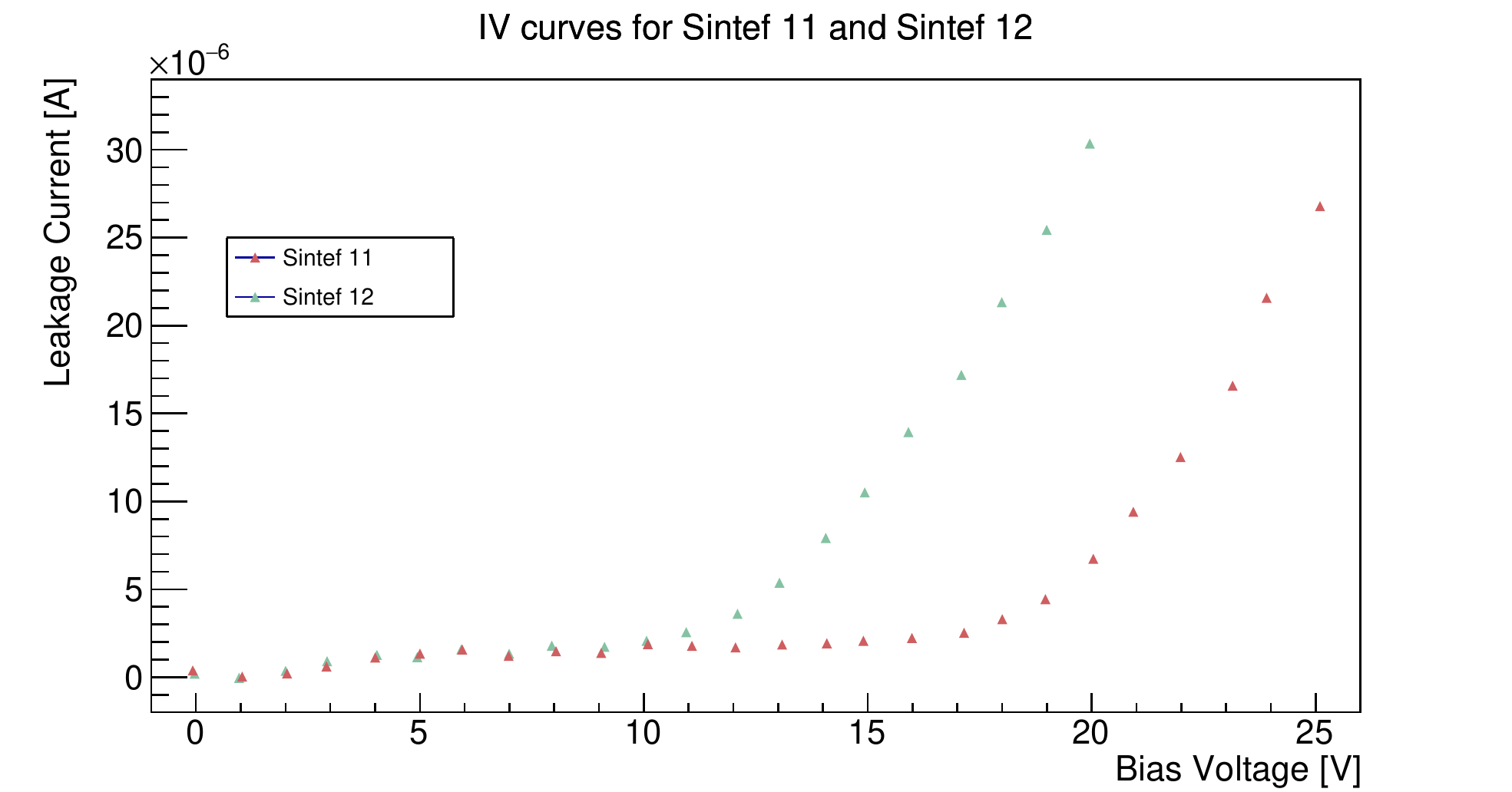}
% "\includegraphics" from the "graphicx" permits to crop (trim+clip)
% and rotate (angle) and image (and much more)
\caption{\label{fig:iv} IV-characteristics of two of the sensors that were mounted in the test beam. }
\end{figure}
\section{Test beam setup and analysis}
 The modules were mounted in a beam line consisting of 120 GeV pions at the CERN H6 testbeam area during two testbeam periods in 2015 and 2016.
The high resolution EUDET Pixel Telescope
\cite{eudet:1} was used to reconstruct the pion tracks through the setup. The telescope consists of
6 pixel planes,  three  placed 12 cm apart upstream of the devices under test (the DUTs) while the remaining three were placed downstream of the DUTs, also 12 cm apart.  The core of the telescope is
the Mimosa26 pixel sensor \cite{eudet:2} with a pitch of 18.5
$\mu$m. Each
plane consisted of 576 x 1152 pixels covering an active area
of 10.6x21.2 mm$^2$. Triggering was achieved by the use of upstream and downstream sets of two 1x2 cm$^2$ scintillators. During the 
second beam period, only five of the six planes were useable in the reconstruction. 

For sufficiently high particle rate, the serial readout of the Mimosa26 sensors could results in ambiguities in the hits associated to a particular event. In many runs it was found that a given trigger contained hits from previous triggers, resulting in several high quality tracks being reconstructed per event. This problem was solved by requiring  that at least one of the FEI4 modules mounted in the setup  could serve as an additional reference.
It was required that a signal in the reference module should be associated to an EUDET track before being passed on for analysis of the DUTs. This resulted in a very clean selection of good events for analysis.  

For the runs studied below, the pointing resolution on the different DUTs was determined to vary from about 4 $\mu$m to about 8 $\mu$m, depending on the position of the DUT in the setup, and on the number of active tracker planes. 
\section{Analysis and results}
Before data collection in the beam, tuning of the FEI4 modules was necessary. The signal height in the FEI4 chip is recorded as a Time over Threshold (ToT) that is digitized in a 4-bit ADC. The modules were tuned 
to give an ADC signal of ToT = 9 as response to a charge release of 20000 electrons. Data were collected at thresholds of 2000, 2500, 3000 and 4000 electrons, and at biases of 5, 7.5, 10 and 15 Volts. When not stated otherwise, the results in the following are for data taken at 10V bias, and a threshold of 3000 electrons. No significant variation of efficiencies were found for the different biases and threshold although dependencies at the level of 1-2 permille could not be ruled out. \cite{andreas}   
  
\subsection{Cluster and signal sizes}
The first step in the data analysis was to assemble all individual hits in a sensor to \emph {clusters}, consisting of a collection of neighbouring pixels with charge deposition. The clustering algorithm added neighboring hits to the cluster until all neighbours with charge deposition were found. Fig \ref{fig:clsiz} shows the distribution of cluster sizes (the number of individual hits included in a cluster), and the distribution of summed ToTs in the clusters. These quantities showed very weak dependence on the settings of the runs (thresholds and biases). 
\begin{figure}[htbp]

\qquad
\includegraphics[width=.4\textwidth]{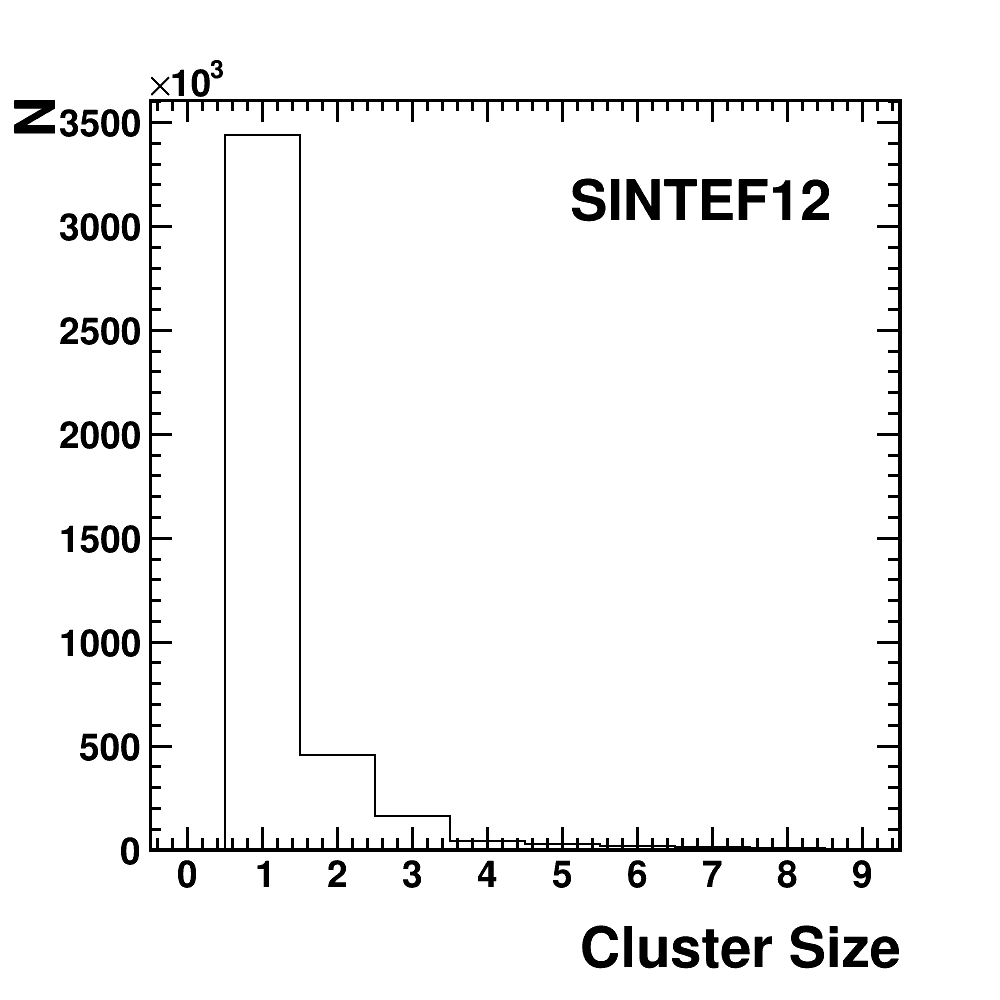}
\includegraphics[width=.4\textwidth]{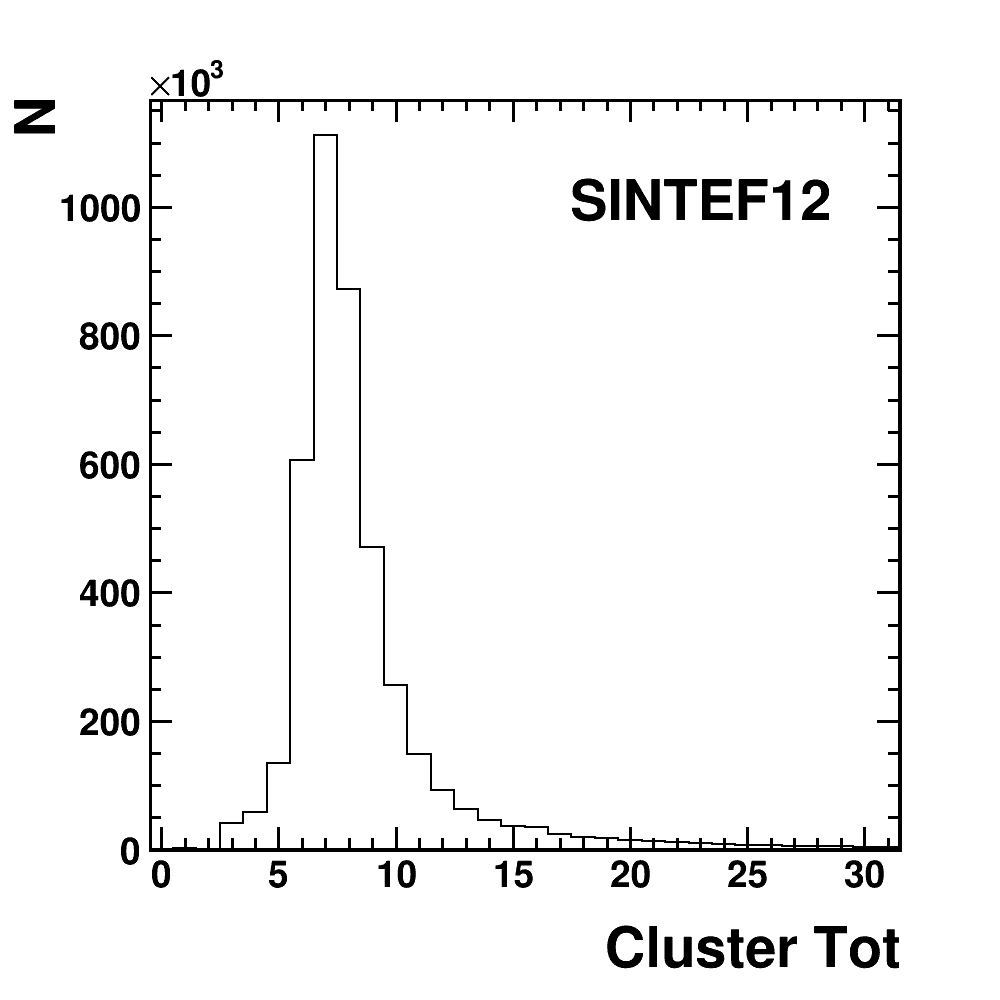}
% "\includegraphics" from the "graphicx" permits to crop (trim+clip)
% and rotate (angle) and image (and much more)
\caption{\label{fig:clsiz} Distribution of cluster sizes (left), and summed times over threshold (right). These data are for the module SINTEF12 with a
bias of 7.5V and a threshold of 2000 electrons. }
\end{figure}
\subsection{Efficiency}
Good tracks, i.e. tracks with low $\chi^2$ and with a hit in the reference FEI4 module associated to it, were used for the efficiency analysis. 
If a signal from a particular DUT could be associated to the track, it was assumed that the signal was caused by the particle producing the track. 
The efficiency was computed as function of the predicted track impact position in the DUT, and is given as the number of times the track could be associated to a pixel divided by the total number of tracks in that impact position.
 The dependence of the efficiency on the window size for track-hit association was studied, and it was found to reach a  slowly rising plateau some 50-100 microns beyond the nominal edges \cite{andreas}. The sensor was defined to have seen the passing particle if a hit pixel was found inside a window of total size $0.25 \times 0.45 \rm{mm}^2 $ around
the  predicted impact position given by the reconstructed track.  

Fig. \ref{fig:effdis} shows the distribution of efficiencies measured individually for each pixel, when the predicted track position is used to give the identity of the pixel. A number of pixels in each sensor had very low efficiency due to unsuccessful bump bonding, and do not show up
in the figure because it falls below the lower limit of 0.5. 

When computing average pixel efficiencies in figures and plots below, only pixels showing up in fig. \ref{fig:effdis} are used (i.e. with efficiencies larger than 0.5). This effectively excluded those areas on the chip that are dead, and clearly included all pixels that are successfully bonded and read out. Subsequent efficiency estimates of active pixels are not  biased by this requirement.
 \begin{figure}[tb]
\centering % \begin{center}/\end{center} takes some additional vertical space
\includegraphics[width=.4\textwidth] {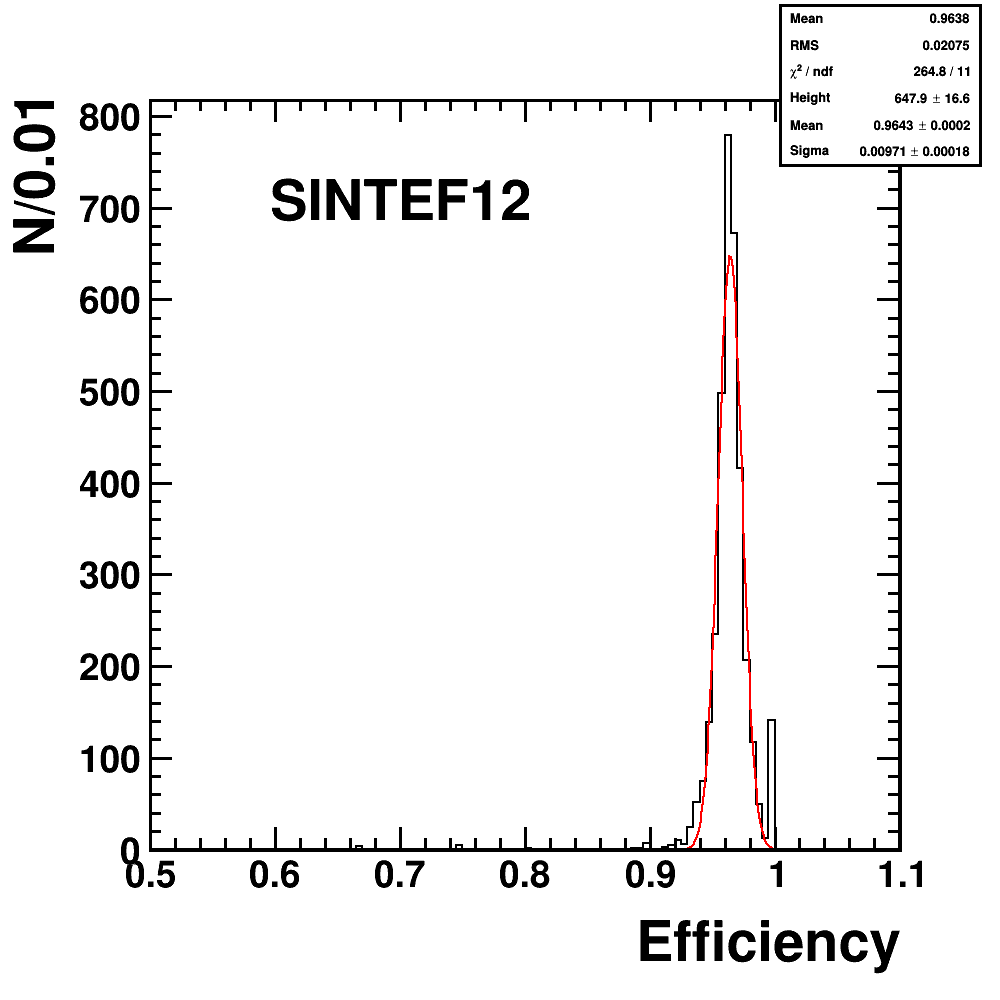}
\includegraphics[width=.4\textwidth] {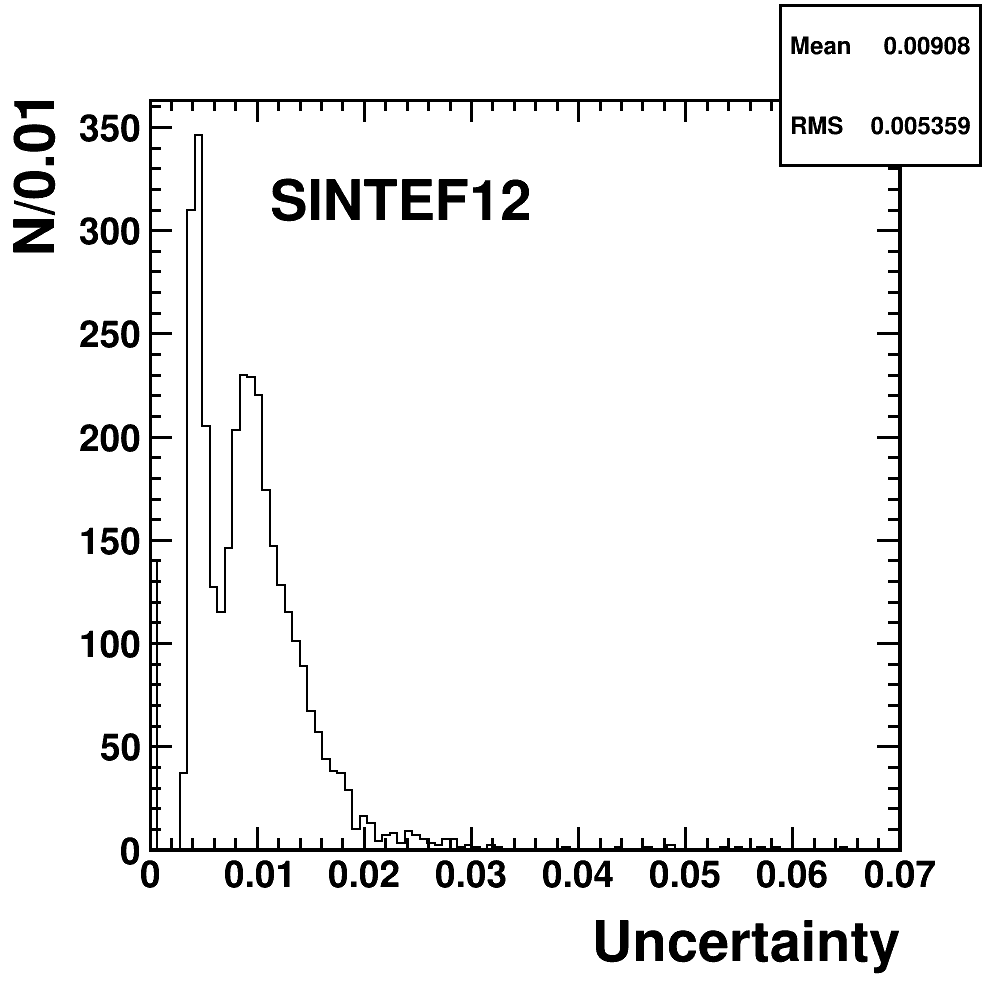}
\caption{\label{fig:effdis} Distribution of individual pixel efficiencies (left), and distribution of statistical uncertainty of the individual efficiency estimates (right). There is variation in the uncertainties due to large differences in the number of tracks passing through the different pixels. The mean value of the estimated uncertainties is close to with the fitted spread in efficiencies observed. This is consistent with expectation when all pixels in the plot have the same true efficiency.}
\end{figure}

The reason for the loss in efficiency in 3D sensors is the presence of the vertical electrodes  in the sensors, in the following denoted by the term 'holes'. In order to demonstrate that this is the case, fig. \ref{fig:eff1} shows the efficiency profile as function of track impact position, averaged over all pixels. It is clearly seen that the loss is indeed at the location of the electrodes. 

In order to characterize this further, the efficiency was computed as function of distance from the nearest nominal electrode center. Fig.  \ref{fig:eff2} shows plots of this efficiency profile. The efficiency 
profile  was modeled as a 
step function smeared with a gaussian, with four parameters; the efficiency in the holes, the efficiency outside the holes, the radius of the holes and the sigma of the the gaussian.  
The fits are reasonable, and the values obtained for the smearing sigma are consistent with the expected tracking resolution (as obtained from studies of single pixel efficiency profiles). The transition region between the holes and the silicon bulk is much smaller than the 
impact resolution, and compatible with a step function. The effective hole radius is found to be close to the nominal 7 
 microns.The values of the fitted parameters are summarized in table \ref{tab:effpars}. However, reliable efficiency estimates in the holes cannot be claimed since the obtained hole efficiencies are not compatible with eachother. Both the step function efficiency modelling as well as gaussian modelling of the pointing resolution are oversimplifications.    
\begin{table}[htbp]
\centering
\caption{\label{tab:effpars} Fitted parameters obtained for the fits
depicted in fig. \ref{fig:eff2}. The quoted uncertainties are statistical only. Note: The fitting procedure for the plateau efficiency returns statistical uncertainties in the fourth digit after the decimal point.}
\smallskip
\begin{tabular}{|l|c|c|c|c|}
\hline
Sensor&Resolution ($\mu$m) &Hole radius ($\mu$m) & Hole efficiency &Plateu efficiency\\

\hline
SINTEF10 & $5.8 \pm 0.2$ & $6.2 \pm 0.5$ & $0.35 \pm 0.04$ & $0.999$ \\
SINTEF11 & $6.2 \pm 0.3 $& $5.2 \pm 0.7 $& $0.32 \pm 0.06$ & $0.998$ \\
SINTEF12 & $3.5 \pm 0.1$ & $6.4 \pm 0.2$ & $0.20 \pm 0.03$ & $0.999 $\\
\hline
\end{tabular}
\end{table} 

The fitted parameters are consistent with expectation:
The spacial track impact resolutions obtained are consistent with what is obtained by other means (studying the smearing of efficiency profiles at pixel edges \cite{andreas}). The hole radii are also consistent with expectation.  Outside the holes, it is seen that the efficiency is extremely high and consistent with 100\% for two of the sensors.  
\begin{figure}[hbtp]
\centering % \begin{center}/\end{center} takes some additional vertical space
\includegraphics[width=.6\textwidth]{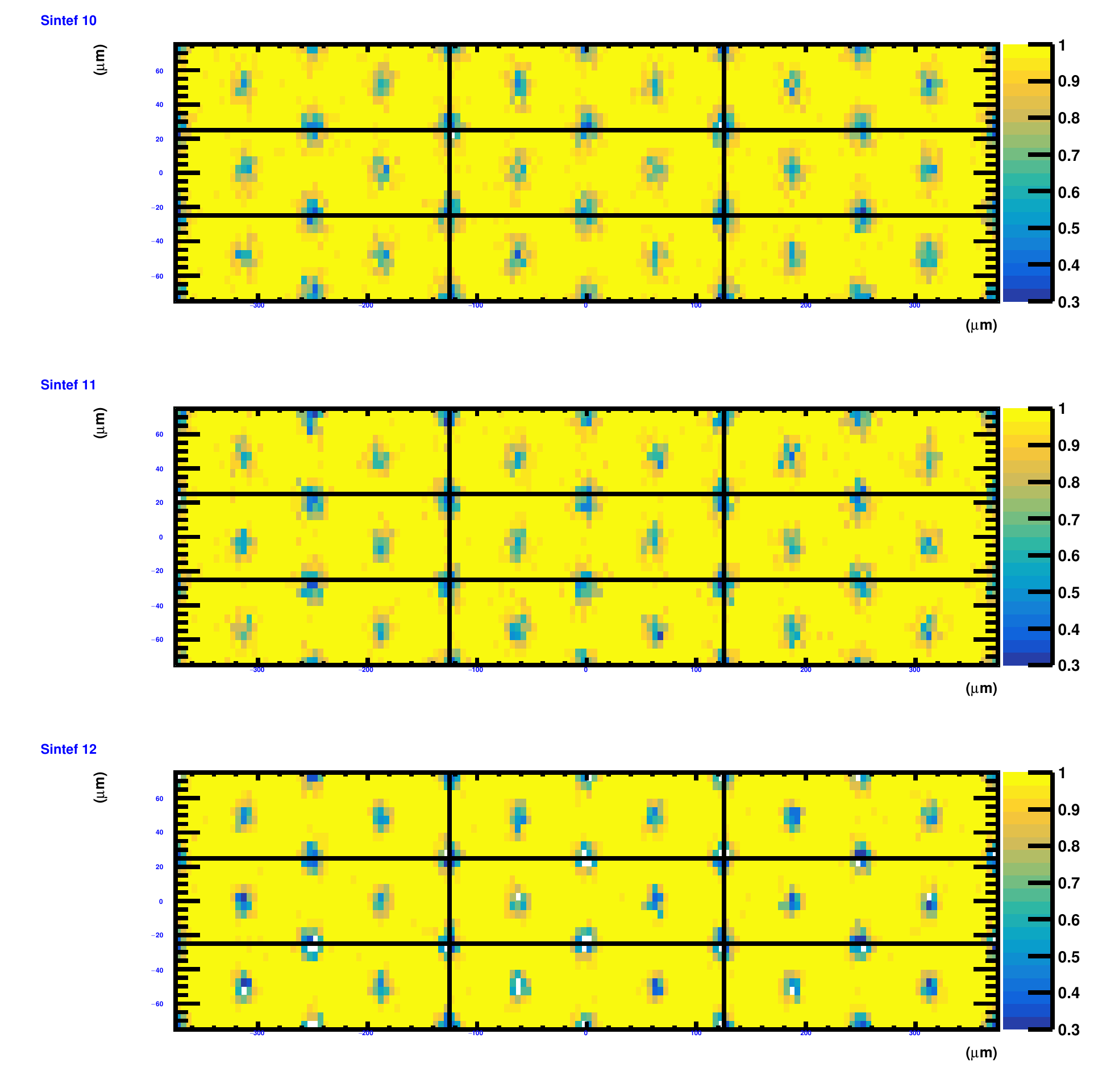}
\caption{\label{fig:eff1} Efficiency maps averaged over the surface of
a pixel area corresponding to $3\times3$ pixels, for the three sensors tested.}
\end{figure}
\begin{figure}[htbp]

\qquad
\includegraphics[width=.9\textwidth]{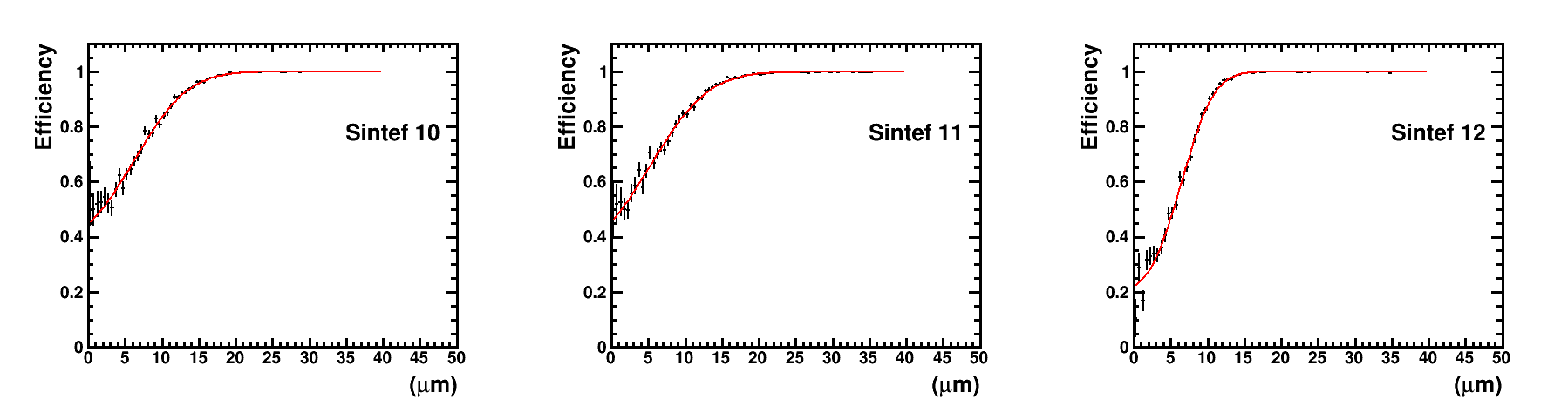}
% "\includegraphics" from the "graphicx" permits to crop (trim+clip)
% and rotate (angle) and image (and much more)
\caption{\label{fig:eff2} Measured efficiencies as function of predicted distance of a track impact from an electrode. The red lines correspond to the efficiency model described in the text. }
\end{figure}

\clearpage
\subsection{Efficiency towards the edges}
In order to evaluate the usefulness of the etched trench around the sensors, special runs were performed where the beam hit the sensor close to the edges. In fig. \ref{fig:edge},
plots of the efficiency  profiles for the outermost pixels along the short and the long edges of the sensors are shown. The sensitive area along the short and long sides are extending about
100 $\mu$m beyond the size of a standard pixel.
\begin{figure}[htbp]
\begin{center}
\includegraphics[width=.5\textwidth]{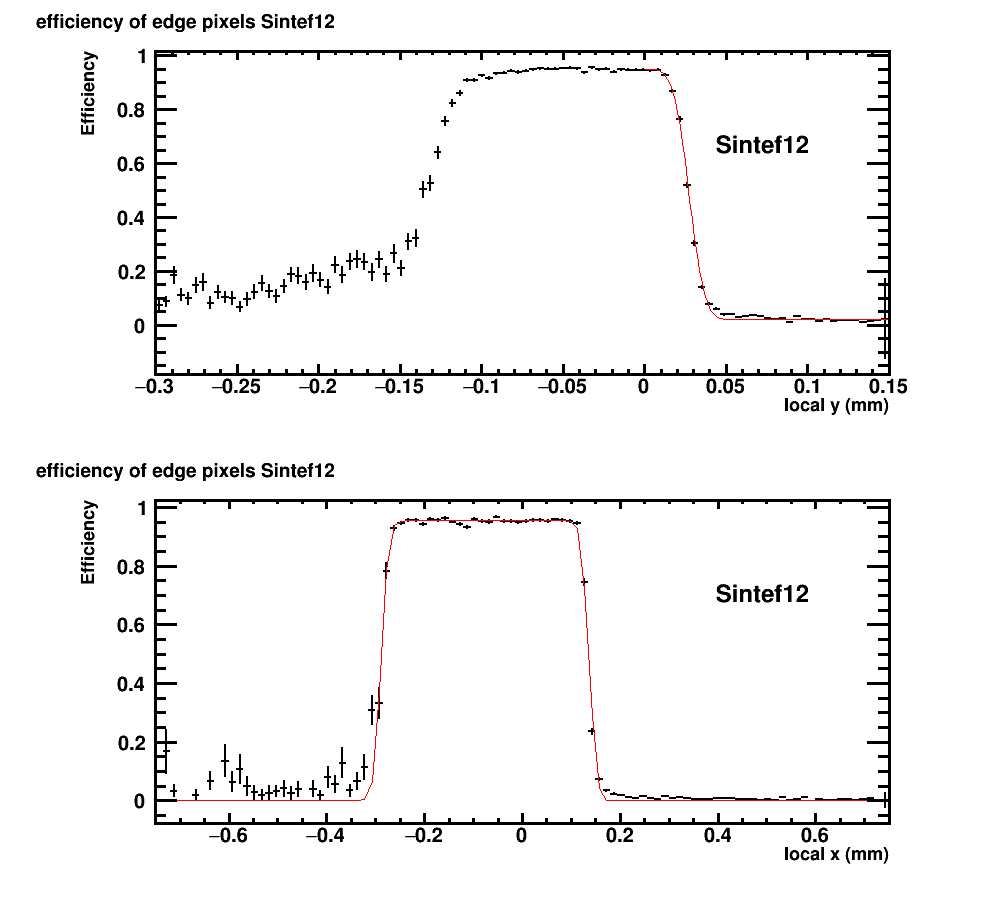}
% "\includegraphics" from the "graphicx" permits to crop (trim+clip)
% and rotate (angle) and image (and much more)
\end{center}
\caption{\label{fig:edge} Efficiency profiles close to the sensor boundaries. The red lines show the result of fits to an 
edge (box) smeared with a gaussian. The upper plot shows the efficiency along the coordinate with 50 $\mu$m pixel pitch. The fitted edge is at y=27 $\mu$m.  The 16 $\mu$m trench is centered at $y=-0.1 $mm. In the lower plot, the fitted width of the box is 425 $\mu$m, centered at x=-76.5 $\mu$m, thus placing the left edge at x=-289 $\mu$m. The trench is location is nominally at $x=-0.2$mm.}
\end{figure}
\section{Conclusion and outlook}
It is shown that 3D sensors produced at  SINTEF are performing as expected with full efficiency at a bias voltages  between 5 and 15V. 
Given the inherent radiation hardness of the design, these sensors might thus be considered as alternatives for use in harsh radiation conditions, such as in experiments at the High luminosity LHC. 
A number of radiation hardness studies of SINTEF 3D sensors from this batch are reported in \cite{Leakage:Laura} and show that properties after irradiation are as expected. Unfortunately, due to the very low number of modules available, complete results on module functionality after proton irradiation could not be extracted for this batch of sensors.  Sensors from a new production at SINTEF have recently become available ("Run 4"). Preliminary wafer measurements show promising results on yield. Work is underway to perform complete tests of sensors from Run 4.   

\section*{Acknowledgements}
The authors wish to thank fellow  testbeam users for technical help and useful discussions during the common data taking efforts.
Furthermore we thank SINTEF MiNaLab for fruitful collaboration. This work is supported by RCN, the Research Council of Norway.



\begin{thebibliography}{99}
\bibitem{intro:1} S.I. Parker, C.J. Kenney and J. Segal,
\emph{3D - A proposed new architecture for solid-state silicon
detectors},
Nucl. Instrum. Meth.
A {\bf 395}
(1997) 328
\bibitem{intro:2}
C. Kenney, S. Parker, J. Segal and C. Storment, 
\emph{Silicon detectors with 3-D electrode arrays:
fabrication and initial test results.}, IEEE Trans. Nucl. Sci. {\bf 46} (1999) 1224.
\bibitem{intro:3} 
T.-E. Hansen et al., \emph{First fabrication of full 3D-detectors at SINTEF}, 2009 JINST {\bf 4} P03010.
\bibitem{intro:4} 
G. Pellegrini et al., \emph{First double-sided 3-D detectors fabricated at CNM-IMB},
Nucl. Instrum. Meth. A {\bf 592} (2008) 38.
\bibitem{intro:5} 
G. Giacomini et al., \emph{Development of double-sided full-passing-column 3D sensors at FBK},
IEEE Trans. Nucl. Sci. {\bf 60} (2013) 2357.

\bibitem{intro:6} J. Lange, \emph{Recent Progress on 3D Silicon Detectors} 24th International Workshop on Vertex Detectors, 1-5 June 2015 Santa Fe, New Mexico. arXiv:1511.02080. 

\bibitem{Sensor:1} 
M. Barbero et al., \emph{ FE-I4,the new ATLAS pixel chip for upgraded LHC luminosities}, Proceedings of
the IEEE Nuclear Science Symposium and Medical Imaging Conference, October 30 November 3,
Knoxville, U.S.A. (2010), ATL-UPGRADE-SLIDE-2009-319.
%\emph{The FEI4 chip guide}. Available at \begin{verbatim} https://indico.cern.ch/event/261840/contributions/1594374/attachments/462649/641213/FE-I4B_V2.3.pdf \end{verbatim} (June 2nd 2018).
\bibitem{Leakage:Laura} L. Franconi: \emph{Insertable B-Layer integration in the ATLAS experiment and development of future 3D silicon pixel sensors}, PhD thesis, University of Oslo (2018), and CERN-THESIS-2018-029.

\bibitem{eudet:1} EUDET Collaboration, D. Haas et al.: 
\emph{A pixelated Telescope for the E.U. Detector R\&D},
Proceedings of the International Linear Collider Workshop
LCWS 2007 and ILC 2007, DESY-PROC-2008-03.  Proc. of the LCWS2007, (2007).
\bibitem{eudet:2}  J.Baudot et al.,\emph{First test results Of MIMOSA-26, a fast CMOS sensor with integrated zero suppression and digitized output} NSS Conference Record
 (2009)   DOI: 10.1109/NSSMIC.2009.5402399.
\bibitem{andreas} A. Heggelund, \emph{Analysis of 3D Pixel Detectors for the ATLAS Inner Tracker Upgrade} MSc. thesis, University of Bergen, Norway (2017). Available through the University of Bergen open research archive (bora.uib.no).



% Please avoid comments such as "For a review'', "For some examples",
% "and references therein" or move them in the text. In general,
% please leave only references in the bibliography and move all
% accessory text in footnotes.

% Also, please have only one work for each \bibitem.


\end{thebibliography}
\end{document}